# Application of PLM processes to respond to mechanical SMEs needs.

Julien Le Duigou[1], Alain Bernard[1], Nicolas Perry[2], Jean-Charles Delplace[3]

[1]IRCCyN, Ecole Centrale de Nantes, 1 rue de la Noë, 44321 Nantes Cedex 03, France
Julien.leduigou@irccyn.ec-nantes.fr, Alain.bernard@irccyn.ec-nantes.fr

[2]LGM[2]B, Université de Bordeaux1, 15 rue Naudet, 33175 Gradignan Cedex, France
Nicolas.perry@iut.u-bordeaux1.fr

[3]Centre technique des industries mécaniques, 57 avenue Félix Louat, 60304 Senlis Cedex, France
Jean-charles.delplace@cetim.fr

**Abstract**

PLM is today a reality for mechanical SMEs. Some companies implement PLM systems very well but others have more difficulties. This paper aims to explain why some SMEs do not success to integrated PLM systems analyzing the needs of mechanical SMEs, the processes to implement to respond to those needs and the actual PLM software functionalities. The proposition of a typology of those companies and the responses of those needs by PLM processes will be explain through the applications of a demonstrator applying appropriate generic data model and modelling framework.

## 1 INTRODUCTION

Due to globalisation, the enterprises have to work in networks more and more diversified and geographically dispersed. To allow this, optimizing the cost, quality and delay, enterprises implement new information and communication technologies. The mechanical SMEs are in the same logical. But those enterprises, even if they are more flexible, have difficulties in front of those new possibilities of exchange and share of information.

PLM systems are one solution to structure and share product information. Nevertheless, in 2007, only 3% of those enterprises have implemented such a system [1, 2]. Our hypothesis is that the functionalities of the actual systems do not feat with the needs of those enterprises.

This paper introduces first the functionalities of PLM systems. Section 3 explains our research approach based on immersions in mechanical SMEs from a PLM dedicated typology. Then the needs of those companies are identified, the processes to respond to those needs are formalised and finally the application of a demonstrator based on a generic model is proposed. Discussion and perspectives conclude this paper.

## 2 PLM PROCESSES

CimData [3] define the PLM as "*a strategic business approach that applies a consistent set of business solutions in support of the collaborative creation, management, dissemination and use of product definition information across the extended enterprise from concept to end of life - integrating people, processes, business system, and information*". The actual definitions of PLM [4, 5] focus on the product data, but do not include the processes management notion. To improve that point, the following definition centred on the activities is proposed:
*"PLM is a business strategy to manage on a collaborative way the product centred activities during the whole lifecycle and across the extended enterprise".*

PLM encompass not only the definition of the product, but also the definition of the product lines, of the technologies used, of the organization, of services associated to the product (services during it use, but also during it maintenance, it end of life…). In a mechanical SMEs context, the production lines are not redefined with the creation of a new product. SMEs have to produce new products with the existing work centres, which is not the case for the firms from automobile or aeronautical sectors where a new production line can be create for the creation of a new product. It is quite the same for technologies used and the organization that do not change with the arrival of a new product. The services are not taken into account by mechanical SMEs, maybe because they are not yet in charge of the maintenance and end of life of their product or because they have to manage the definition of their product before manage their maintenance or their end of life.

PLM is support by business software. The PLM systems help to pass the product definition information from one business solution to another (Figure 1). Interoperability and modularity are important issues in those systems [6]. It encapsulates and then diffuses the information for the product development. It is realized by links between the PLM systems and the other software (Figure 1), as CAD, CAM, ERP, SCM… It ensures the traceability [5], the archiving [7] and the reuse [8] of information.

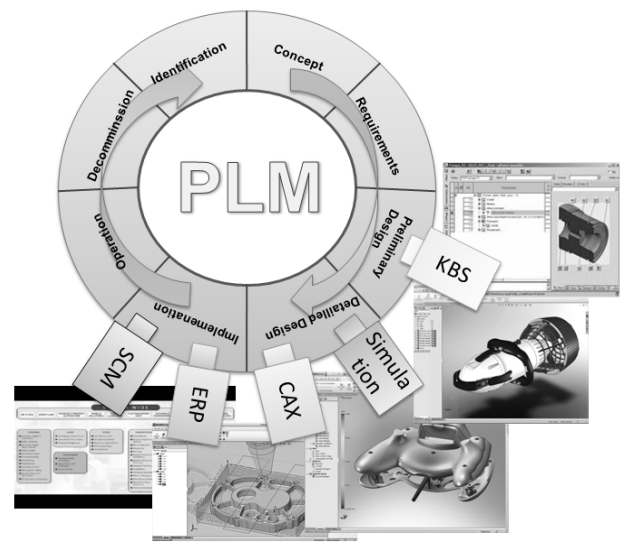

Figure 1: PLM and business software

The PLM systems offer numerous functionalities, the most classical are:

*Vault*: the data and documents are stocked in a server. The data are stocked in a database and structured through the data model implement in the database.

*Right access*: the user must have sufficient right access to operate on the data (read, modify, validate…).

*Check in / check out*: the user must check out the document before modified it. Then he checks in the modified document. This assures that two users cannot modify a document at the same time.

*Versioning*: it traces the change of a document or a data. At a second level, a revision serves to trace major modifications.

*Notification*: notification is the sending of an email with a link to the data or document subject of the email. Automatic notification for change, validation… is called subscription.

*Workflow*: this system simulates processes and automates some activities of the process. It is principally use for documentation processes like validation.

*State*: they are associated to document and define their maturity level (creation, in validation, validated, obsolete).

Actual tools are mainly dedicated to documentation and communication management. The possibility to automate and manage activities more business oriented is not in the actual functionalities. SMEs are looking for more advanced business functionalities.

Notice that the majority of the works done on PLM are related to firm, often automobile or aeronautic assemblers [9, 10, 11]. Deductive approaches are used to integrate SMEs in the assemblers' point of view. An inductive approach will be more appropriated to include SMEs needs.

## 3 RESEARCH APPROACH

To propose processes and functionalities appropriate to SMEs to structure and to exchange product information, our approach inverse the traditional V cycle. It starts with a first inductive and raising step to formalise the needs and the processes to implement. The second step is deductive and downing to validate the propositions (Figure 2).

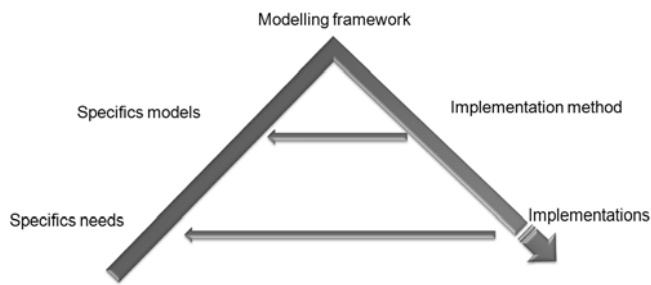

Figure 2: Research approach

From the diversity of the mechanical SMEs, this section proposed a typology to class those enterprises in function of their PLM problematic.

The main technical information link to the product development to manage at an enterprise level is in the Bill of Materials (BoM) and in the routings. The predominant use of one or the other intermediary object depends of the number of components per products of the enterprise.

Indeed the enterprises of which the products have a numerous components work on the assembly of sub-systems. The product information are then in the BoM. For the enterprises of which the products have a few components work on the manufacturing of the parts. The product information are then in the routings.

So the proposed typology has as differentiation axis the number of components per product. The different companies are grouped according to the number of components of the products that they design and/or produce (Figure 3).

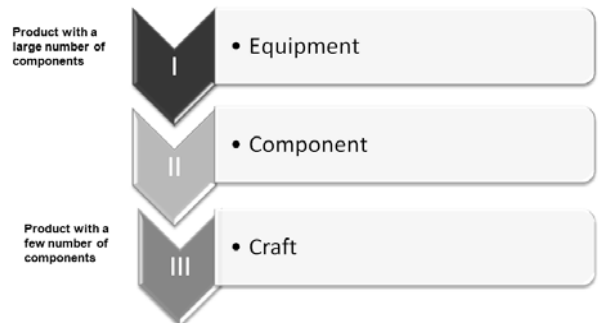

Figure 3: Typology of mechanical SMEs

Three types of companies are obtained:
- The companies producing equipments.
- The companies producing components.
- The companies producing craft parts.

The equipment manufacturers are companies of which the products contain numerous components. They produce special machines, agricultural machines, stevedoring machines… they have problematic about BoM management and traceability, well known of the PLM systems because they match to the needs of the big firms of the automobile and aeronautic sectors which introduce those software. The solutions to implement are lighter and less complex than the big firms' solutions.

The component manufacturers are companies of which the product contains a medium number of components. These enterprises design and/or produce compressors and motors, tools, transmissions (hydraulic, mechanic or pneumatic)… They have various needs from the customer needs analysis to the manufacturing.

The craft parts manufacturers have a few components in their products. They produce discrete parts from the forging, foundry, machining, drawing… these enterprises have problematic about the routings management.

After have apply the approach on three different enterprises, the result is three needs' maps, three processes models and three data models [12, 13]. The aggregation of those results permits to define a generic model for the mechanical SMEs (Figure 4).

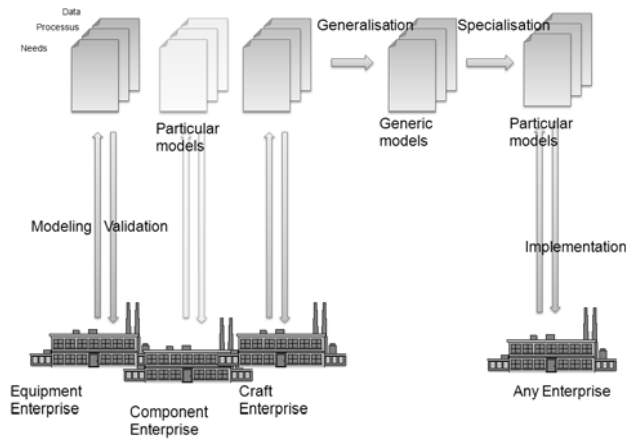

Figure 4: Generalisation approach

## 4 APPLICATION OF PLM PROCESSES

### 4.1 Identification of the needs

The principal problems of each enterprise were identified in [12, 13]. A list of PLM needs is created from those immersions. The needs retain in the list are those which are not fully implement in the actual PLM systems.

Following our study, six classes of needs are identified:

*Configuration management*: it comes from the equipment part manufacturers. The PLM systems must manage the alternative, the options, the versions, the families and the differentiation between internal and external products.

*Collaboration*: the exchange with the customers (for the parts and components manufacturers) and with the suppliers (for the components and equipments manufacturers) must be facilitated and standardized, especially for the CAD files exchanges.

*Multiple views*: the information has to be visible with the structure and the names of each department. It is particularly need for the BoM in design and production departments, corresponding to the structures of the CAD and the ERP.

*Routing management*: the raw parts manufacturers need that the routings and the whole information that it includes (operations, work centres, tools…) could be manage from the PLM systems.

*Interoperability*: the interoperability with the ERP and the CAD is asked from the three types of SMEs, especially for the BoM and routing update.

*Decision aid indicators*: the cost is the indicator the most asked, to choice between alternative products for the equipments manufacturers or to choice between alternative operations in a routing for the raw parts manufacturers.

Those needs have to be resolve by the PLM systems to interest the mechanical SMEs. The processes implement in the pilot enterprises solve those needs. The generalisation of those processes should help to solve the generic needs.

### 4.2 Formalisation of the processes

The processes covert by the pilot companies could be aggregate to represent a complete development product in an extended enterprise.

The required level of detail of the SADT makes appear the intermediaries. To know the processes of creation, use and decommission of those intermediaries in the information system, it is necessary to know those processes in the real life with the same level of detail.

Moreover it is necessary to reach this level of detail in the three companies to be able to generalize appropriately the processes. Indeed the same intermediaries are in several companies, continuing their life cycle passing from a company to another. This exchange of intermediaries allows linking the processes from one company to another, like in an extended enterprise.

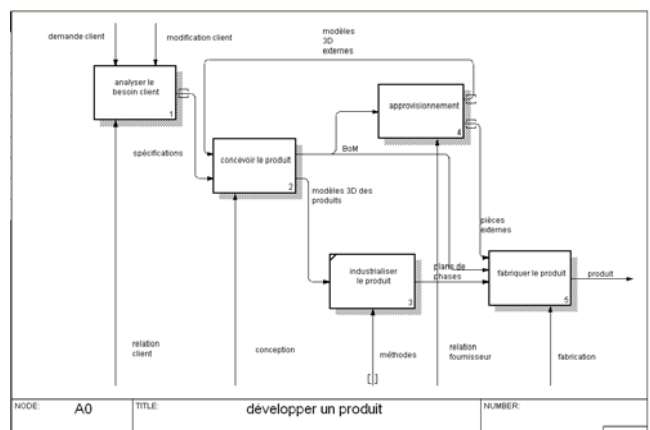

Figure 5: PLM processes

The global process Figure 5 is a development process. It includes the need analysis, the design and the industrialization. The production, the sale, the distribution and the end of life of the product are not covert by this process which is focus on the SMEs context. As explain previously, the mechanical SMEs do not express the wish to include those processes.

To manage those processes, an appropriate data model is required. The objects to include in the data model are those used by the processes and which appear as input and output of the activities.

### 4.3 Realisation of the processes

The model generalizes the objects used in the pilot companies to obtain generic objects. The semantic alignment was the principal difficulty. One object can be named differently depending of the enterprise. The fact that one consultant have done the three models makes easier the concatenation of the objects. The generic objects are represented in the UML class diagram Figure 6.

The different identified objects are the activity, the product, the material, the function, the resource, the trigger, the organization and the document. The activity specializes into task and operation, the organization into customer and supplier and the resource into material, human and software resource. More explanations about the generic model are in [14].

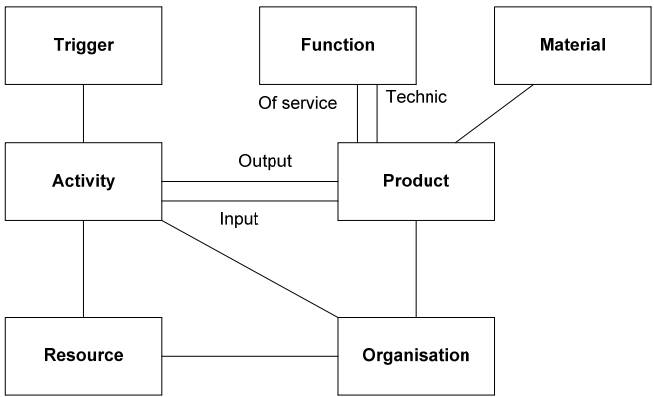

Figure 6: Class diagram of the generic model

The objects at the generic level are not enough specific to be really useful in a particular company. A framework that proposed an instantiation of the generic model is proposed to match with the particular objects of each enterprise keeping a high level semantic alignment.

Based on modelling approaches like CIMOSA or GERAM [15, 16], the framework proposes a system with needs, processes and data model coherent for the three pilot enterprises. Figure 7 shows a cube representing the framework. The abscissa represents the three results of the modelling method: the needs map, the processes model and the data model. The ordinate represents the instantiation of the three components, the generic, partial or particular levels. And the depth represents the different views of the systems, process, information, resource and organization.

The generic model proposes a union of the views process, information and resource. It also integrated the organizational view that coordinates the other views. This axe notices the different views that are modelled, even if they are all integrated in the generic model.

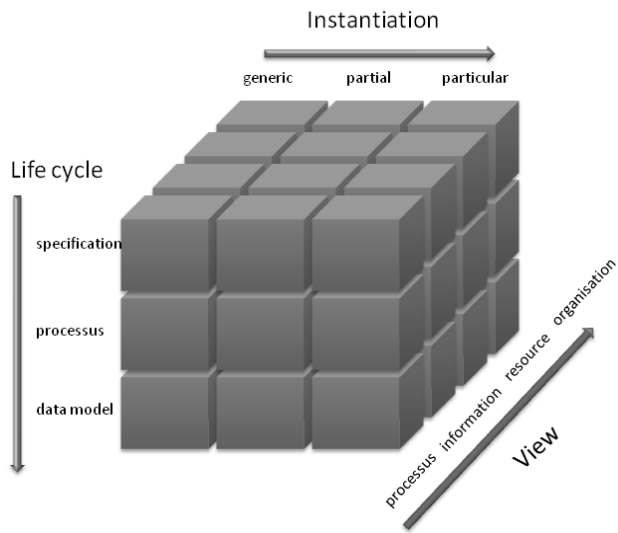

Figure 7: Modelling framework

A software demonstrator is created to automate the use of the framework in an industrial context. The architecture of the demonstrator is a rich client/ server (Figure 8). The client is developed in VB.Net and the server uses MS SQL Server. They communicate through SQL requests via http/https.

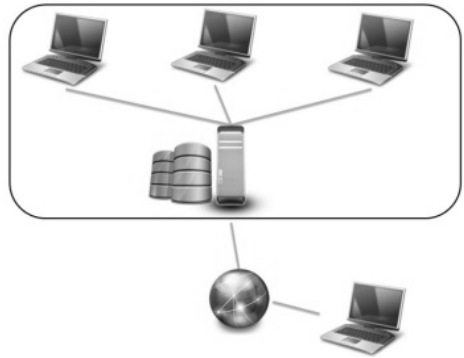

Figure 8: Demonstrator's architecture

The proposed generic model is implemented in the demonstrator, with the generic classes, attributes and methods. Each class is specialisable in sub-class, inheriting of the attributes and methods of the father class. New attributes could be added to the new class. A new specialization is applicable to the sub-classes and so on.

The interface is constituted of a tool bar, a treeview, a fileview, a data card and a viewer (Figure 9).

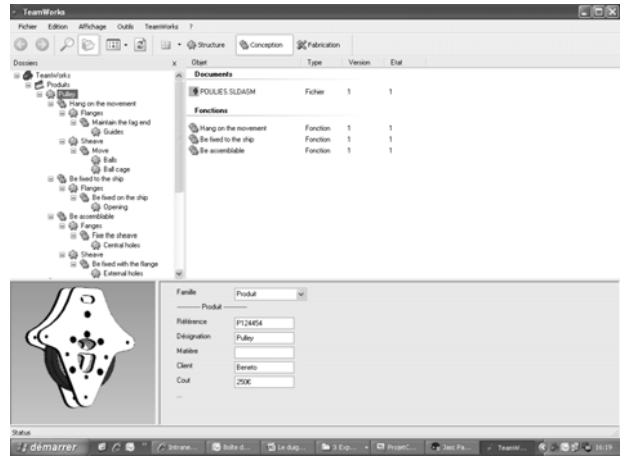

Figure 9: Demonstrator's interface

## 5 APPLICATION OF THE DEMONSTRATOR

The application is chosen to cover the maximum of identified needs. The study needs are: the configuration management, the link with the supplier, the multiple views, the link with the CAD and ERP, the routings management and the cost indicator.

The configuration management is partially covered by actual PLM systems. It is principally a request from the equipment manufacturer. The creation of product family and the differentiation between internal and external products is nevertheless simplified in the demonstrator.

To create a product family, the user right clicks on the generic class of product on the treeview and select new. A new sub-class (a family of product) is created, inheriting of the attributes of the product class plus the new that the user defines for this family. Each product create in this family will obtain the attribute of the family. Sub family can be created reproducing the same method on a family.

The internal and external products are in the referential of the enterprise with different colours (yellow for the internal products and green for the external products). The external product can be taken directly from the supplier PLM systems. The different suppliers identified are present in a specific folder. A drag and drop from a component of the supplier to the demonstrator imports the product, all the attributes and the attached files.

The composition of a product is not the same depending of the user's work in the enterprise. There is no universal view appropriate for all the business units and each business view gives advantages to their users. In the present work, three views are retained: functional, structural and manufacturing [17].

In the structure view (Figure 10), the components of the product are decomposed in the treeview. To obtain a functional view, the user has to change the view clicking on the function button. Then he can create functions under the product (the internal functions) and add in the components of the product (the functions of service of the components). So the product is decomposed from a functional point of view.

In the manufacturing view, the assembling operations are linking to the product and to the component (Figure 11). It gives a manufacturing structure of the product through a product – assembly – component decomposition of the product.

In each view, some objects can be added separately. The object is visible in the selected view and not in the others. It could be the case for the addition of grease in the manufacturing BoM that do not have to be visible in the design BoM.

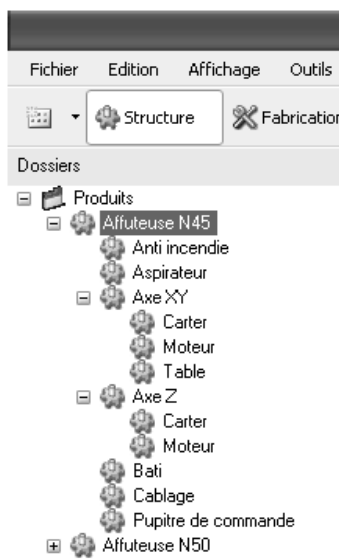

Figure 10: Structural decomposition of a grinding machine

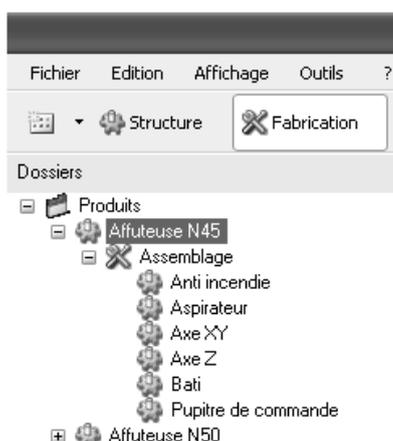

Figure 11: Manufacturing decomposition of a grinding machine

The interoperability of the PLM systems with the CAD and the ERP are the most important link to create [18].

For the CAD link, the designer imports the assembly treeview of the product with a drag and drop of the CAD file from the office to the demonstrator. The CAD treeview is then copied in the structure view. The components are created and the corresponding CAD files are imported in the right component objects. With the import of the CAD file, a link is created between the CAD attributes and the object attributes (for the attributes with the same name in the CAD system and in the demonstrator). Those attributes are then synchronized. The implementation is done on SolidWorks from Dassault Systems.

The manufacturing view can be synchronised with the ERP of the enterprise to obtain BoM and routings up to date. A mapping with an ERP is implemented in the demonstrator. It synchronised the items, the operations, the work centres, and all their respective attributes, updating the BoM and the routings. This is done on the ERP EFACS from EXEL Computer System, using the data base Informix from IBM.

The management of the routings directly in the PLM systems is a wish of the components and raw parts manufacturers.

To create a routing, the user creates the different operations in the product (Figure 12). He can define optional operations, alternative operations and external operations (appearing in green and not in yellow as the internal operations). Then he loads the resources useful to each operation dropping the input items, the work centres, the tools…

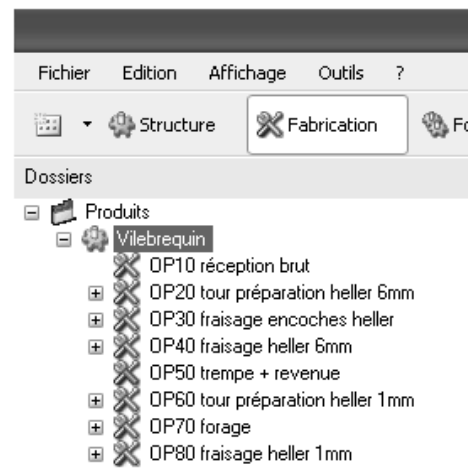

Figure 12: Routing of a crankshaft

The indicator the most requests for aid design and industrialisation in SMEs is the cost of the product, including the cost of the manufacturing processes.

The cost notion is present in the demonstrator. To have an estimation of the cost of a product, the user must create the routing of its product. It includes the components, the work centres and all the other consumables. Then the cost of each operation, each work centres and each component of the lower level of his decomposition is filled. Then the demonstrator calculates the estimated cost of the product.

## 6 DISCUSSION

The contribution of this work is in two parts: in one hand the identification of the needs of mechanical SMEs in terms of PLM, and in the other the resolution of those needs based on a generic data model.

The identification of the needs of mechanical SMEs is done based on three representative companies. The number of studied companies is few for a generalisation but we assume that the choice of the companies based on the typology explain in section 3, combining with the immersion in those companies gives a realistic view of the problems encounter by the mechanical SMEs on this subject.

The second point is the realisation of the processes based on the generic model and adapted to the particular company following a modelling framework. The demonstrator implements the objects of the generic model and its specialisation. The implementation of this demonstrator in an enterprise on a major project will validated the robustness and the appropriateness of our approach.

## 7 CONCLUSIONS AND PERSPECTIVES

The SMEs do not integrated easily the PLM systems. Their needs do not match to the actual PLM functionalities. Our approach classifies the needs of mechanical SMEs based on immersions in representatives companies. Then the processes to be implemented in those enterprises to respond to those needs are formalized. And then the data model to implement in the PLM system is deduced from the processes. Finally the data model is implement in a demonstrator and it use to solves identified needs showing the technical feasibility and the applicability of our approach.

The next contribution will be a detailed explanation of the generic model and the modeling framework to allow its use by other companies.

## 8 ACKNOWLEDGEMENTS

We would like to thanks PSL CONCEPT, CAPRICORN and SMP companies for allowing us to carry out our study in their companies and for their technical support.